Manuscript Title: Negative Spillover: A Potential Source of Bias in Pragmatic Clinical Trials

Author Name: Sean Mann, M.Sc.

Affiliation: RAND Corporation, Santa Monica, California, United States of America

Contact Information: smann@rand.org

Declaration of Competing Interests: No conflicts of interest exist.

Acknowledgements: The author would like to thank Beth Ann Griffin, Edward N. Okeke, Carl Berdahl, Maria DeYoreo, Kerry Reynolds, Kortney Floyd James, Claude Messan Setodji, Zachary Predmore, and Mayda Nathan for their helpful feedback on this manuscript.

Abstract: Pragmatic clinical trials evaluate the effectiveness of health interventions in real-world settings. Negative spillover can arise in a pragmatic trial if the study intervention affects how scarce resources are allocated between patients in the intervention and comparison groups. This can harm patients assigned to the control group and lead to overestimation of treatment effect. While this type of negative spillover is often addressed in trials of social welfare and public health interventions, there is little recognition of this source of bias in the medical literature. In this article, I examine what causes negative spillover and how it may have led clinical trial investigators to overestimate the effect of patient navigation, AI-based physiological alarms, and elective induction of labor. I also suggest ways to detect negative spillover and design trials that avoid this potential source of bias.

Pragmatic clinical trials are used to evaluate the effectiveness of health interventions in real-world settings. In these trials, patients are randomly assigned to receive either usual care or an intervention.[1] Randomization serves to produce balanced groups of patients, so that differences in study outcomes can be attributed solely to the effect of the intervention.[2] Randomized trials are considered the "gold standard" of evidence and their results can lead to changes in clinical practice.[3,4] Yet one aspect of these trials' real-world settings is often overlooked and could lead to biased results.

Clinical resources in real-world settings are sometimes scarce, meaning that their supply is insufficient to meet the needs of all patients.[5-8] During a trial, if an intervention increases patients' utilization of scarce medical resources (such as specialist appointments or hospital beds) this can negatively affect availability of care for patients in the control group. This problem is referred to by economists as "negative spillover" or "crowding out" and is sometimes addressed in trials of social welfare and public health interventions.[9-11] However, there is little recognition of this potential source of bias in the medical literature. As a result, some clinical trials may have unknowingly overestimated the effect of a diverse set of health interventions.

For example, clinical trials evaluating patient navigation have randomly assigned patients to receive help obtaining clinical appointments. One such trial found that navigation led these patients to receive earlier diagnoses than comparison patients assigned to usual care.[12] However, the supply of appointments in the trial setting was likely constrained; navigations' effect may have been due to its directing scarce appointment slots towards the treatment group at the expense of the comparison group. If so, navigation might have no effect if provided equally to all eligible patients, even in the same health system in which the trial was conducted.

This problem resembles another better known source of bias: lack of blinding can also result in different levels of care provided to treatment and comparison groups.[13,14] Yet even a study that is double-blinded, in which trial subjects and clinicians are unaware of group assignment, could suffer from spillover if an intervention, by its very nature, affects use of a scarce resource shared across study arms.

In this article, I examine how sharing scarce resources can lead to negative spillover and how it might affect outcomes in trials of patient navigation, elective induction of labor, and physiological alarms. I also suggest ways to detect negative spillover and design trials that avoid this potential source of bias.

**Sharing Scarce Resources Can Lead to Negative Spillover**

Four conditions are necessary for scarcity-related negative spillover to affect trial results, as shown in Figure 1. Each of these conditions concerns a different aspect of a study: design, setting, intervention, and outcomes.

*Condition 1. Resources are Shared by Intervention and Comparison Groups*

Group assignment is a key aspect of trial design. In many trials, individual patients are randomly assigned to either the intervention or comparison group. When these patients are all served by the same clinicians or hospital, the care they receive draws upon a shared pool of medical resources.

*Condition 2. A Scarce Resource is Present in the Trial Setting*

Scarcity is most apparent during health emergencies or in cases of "chronically limited" tangible items such as organs for transplantation.[5,6,8] Scarcity is less visible when it concerns an intangible resource and is an accepted part of clinical practice.[5]

Clinician availability is a primary example of a resource whose scarcity is often overlooked.[5,7] In primary care, intensive care, and other health settings, clinicians must routinely decide how much time to spend caring for one patient rather than another.[7,15-17]

*Condition 3. The Study Intervention Affects Utilization of the Scarce Resource*

Pragmatic trials often study the effectiveness of interventions that influence how care is delivered to patients.[1,18] The intervention under investigation might be a new health information technology, care coordination process, or medical procedure. Compared to usual care, these interventions can vary in their utilization of clinical resources. A new surgical procedure might take longer and thus require greater use of an operating room. A new approach to managing chronic conditions might involve more visits to specialists. Study interventions may affect demand for care resources in multiple and sometimes unanticipated ways.

*Condition 4. Availability of the Scarce Resource Affects Study Outcomes*

The availability of resources such as clinical staff or intensive care unit (ICU) beds can affect patient outcomes such as mortality, satisfaction with care, hospital length of stay, or the time it takes to receive a definitive diagnosis.[19-24] Such outcomes, which are important to patients and relevant to clinical decision-making, often serve as study endpoints in pragmatic trials.[1,18]

**Negative Spillover May Have Affected a Diverse Set of Studies**

Trials evaluating a diverse set of health interventions may have met the four conditions that lead to negative spillover, as shown in Table 1. This table is a partial list only; it is provided to illustrate how a range of trials may have been affected and is not the result of a systematic search or literature review.

*Patient Navigation*

Patient navigation for cancer diagnosis is designed to help patients schedule and receive diagnostic services following abnormal screening results. One randomized trial reported that navigation helped patients obtain diagnostic resolution earlier and at higher rates for breast and colorectal cancers.[12] Another randomized trial reported similar results for breast cancer, and additionally found that patients without cancer assigned to navigation received faster diagnostic resolution than patients with cancer assigned to usual care.[25] Both trials have been held up as among the best evidence in support of patient navigation in systematic reviews.[26-28] These reviews were subsequently cited by clinical practice guidelines to support patient navigation for colorectal and breast cancer services.[29-31]

In both trials, patients in the intervention and control arms received care in the same hospital or health system. Neither study reported on availability of diagnostic specialists during the trial. Yet concerns over appointment availability for cancer diagnostic services are common[21] and wait times may lengthen due to increased demand[32,33] or fewer available staff.[34] Navigation might have helped patients obtain more appointments with specialists at earlier dates; this would be consistent with the trials' findings. Greater utilization of appointments by patients assisted by navigators could have reduced availability of appointments for others and led to delays in diagnostic resolution for patients assigned to usual care.

*Induction of Labor*

A multicenter trial randomly assigned low-risk pregnant women to labor induction at 39 weeks or to usual care in the form of expectant management. The study found that induction of labor did not increase the frequency of adverse perinatal outcomes.[35] The American College of Obstetricians and Gynecologists issued a clinical practice advisory stating that "based on the findings demonstrated in this trial, it is reasonable for obstetricians and health-care facilities to offer elective induction of labor to low-risk nulliparous women at 39 weeks gestation."[36]

During the trial, patients in the induction and expectant management groups received care in the same hospitals. Labor and delivery rooms were not always available, which affected some patients' care. Patients in the induction group stayed an average of 45 percent longer in the labor and delivery unit which could have reduced clinicians' availability to care for patients assigned to expectant management. Given evidence that higher clinician workload negatively affects perinatal outcomes,[37-39] negative spillover could have biased trial findings.

*Physiological Alert Systems*

Current guidance on sepsis management notes growing interest in algorithm-based physiological alert systems that could support timely treatment of sepsis.[40,41] One study cited in this guidance randomly assigned ICU patients to monitoring by the usual sepsis detection system or the usual system plus a machine learning-based alert system.[42] The trial reported that in-hospital mortality decreased by over 50% for patients monitored by the additional alert system.

Alerts from the additional system may have directed extra clinician attention to patients in the intervention group. Clinician time and attention is often limited in ICU settings, which can lead to rationing of care and decreased adherence to infection-control protocols.[5,16,43-46] Exposure to excessive numbers of alarms can overload clinicians and contribute to missed alarms and patient deaths.[47,48] If patients in the intervention group received extra attention at the expense of clinicians' availability to care for control patients, negative spillover could have occurred.

*Other Interventions and Study Designs*

Table 1 also contains studies of other interventions that may have influenced allocation of scarce resources between study arms. One randomized trial evaluated the effect of text-based reminders on COVID-19 vaccine uptake in early 2021 when vaccines were not widely available.[49] Another trial evaluated the impact of a new genomic screening test on time to diagnosis as compared to standard tests conducted at the same laboratory;[50] if technicians' workload increased due to the new test this could have delayed test results for all patients. A third trial evaluated the effect of an automated reminder system on timely follow-up of abnormal test results, though all patients were in the same two health systems and their care may have drawn from a shared pool of diagnostic resources.[51]

Pragmatic trials are not the only type of study that might be affected by negative spillover. For example, cohort studies have also been used to examine the effect of patient navigation programs.[52,53] In these studies, if navigation affected allocation of a scarce resource across cohorts, then spillover may have occurred. The conditions that lead to negative spillover could also be present in an explanatory randomized controlled trial, though this may be less likely due to efforts to ensure such trials are held under "ideal," rather than real-world, conditions.[1,18]

**Implications**

Negative spillover is not just a source of bias. It also has implications for patient safety. Researchers should not simply assume that a clinical trial will do no harm to patients assigned to usual care. If negative spillover occurs, these patients are no longer receiving "usual care" as it is commonly understood. Instead, they may be receiving substandard care, if clinical resources they need have been reallocated to others via random assignment. In addition, when a study intervention affects allocation of a scarce resource, the welfare of patients who share that resource but are not direct trial participants should also be considered.[54,55] Clinicians can also be harmed if an intervention increases workload and contributes to burnout or moral distress.[19,55,56]

Spillover due to sharing of scarce resources will not uniformly lead to overestimation of treatment effect and harm to patients in the comparison arm; spillover could also occur in a positive direction. Patient navigation may result in lower utilization of diagnostic resources due to fewer missed appointments. Elective induction of labor might be rescheduled to reduce clinician workload when a labor and delivery unit is crowded. A physiological alarm system that results in timely treatment could reduce overall patient acuity in an ICU. Such mechanisms could free up medical resources and thus improve outcomes for patients assigned to a comparison group, leading investigators to underestimate the beneficial effects of an intervention.

Study results that may have been affected by negative spillover can still be a valid indication that an intervention affects care utilization. For example, patient navigation for underserved populations could still be used to decrease disparities in access to diagnostic services, even if navigation simply reallocates scarce appointment slots to these groups.

*Detecting Negative Spillover*

It is not possible to firmly conclude whether, or to what extent, negative spillover occurred in the trials discussed here. In some cases, study data could be reanalyzed for signs of spillover. If spillover occurred, patients assigned to usual care during a trial could have experienced changes in outcomes when compared to a pre-trial baseline. For example, prior to the trial evaluating induction of labor, study investigators estimated an expected risk-adjusted rate of adverse perinatal events at 3.5% for the expectant management group based on rates previously observed in the study setting.[35] During the trial, however, these patients experienced adverse events at a higher rate (5.4%) which could be a sign that they faced reduced access to care due to higher resource utilization by patients in the induction group.

Researchers could also compare subgroup outcomes to test the hypothesis that a specific type of patient, study site, or time period included in a trial were more affected by cross-arm sharing of scarce resources. Data on resource availability and utilization would be especially useful in such analysis. Ultimately, however, obtaining definitive evidence on negative spillover might require conducting a two-level randomized trial like those used to identify other spillover effects.[57,58]

*Designing Spillover-Proof Trials*

Investigators, research funders, institutional review boards, health system managers, patient organizations, and other stakeholders[54,59] should consider whether the conditions that lead to negative spillover might be present in any proposed trial. Intangible resources, such as clinicians' time and attention, should be a particular focus given that they are easily overlooked. If a study intervention might affect allocation of a possibly scarce clinical resource, then the trial should be designed to avoid spillover.

Cluster-randomized trial designs can avoid negative spillover if the unit of randomization (such as a clinical practice or hospital) contains its own distinct pool of resources used to care for trial participants. Cluster randomization might not always be feasible, however, as it often requires higher numbers of participants and coordination across multiple study sites.[2]

In some cases, a trial could be designed to balance resource availability or demand across intervention and control groups within a single site, as shown in Figure 2. The proposed trial designs share characteristics with attention-control study designs used to limit bias in trials of behavioral interventions.[60] For example, if an intervention involves use of discrete resources such as vaccination appointments to achieve an outcome, a trial might randomize both patients and appointment slots, in the same proportions, to the intervention and comparison arms. Similarly, an intervention that changes how patient needs are translated into a demand signal for care, such as a new physiological alarm system, might be calibrated to generate the same average number of alarms per patient as the existing system that it is compared to. Where practical, investigators might use such demand- or resource-balanced trial designs to mitigate negative spillover while also avoiding the larger costs associated with cluster randomized designs.

Finally, if trial design cannot be changed, researchers might still collect data on resource availability and utilization (ideally at baseline and during the trial) to help identify and control for potential spillover. Common data models intended to support comparative effectiveness research[61,62] might also consider adding elements that measure availability of potentially scarce resources in study settings.

*Improving Methods Guidance*

Guidance on clinical trial methods does not currently address the possibility of negative spillover. This includes trial design,[1,2,4,63] reporting,[64-66] and program guidance[67,68] as well as frameworks used to identify potential harms to research participants[69-71] or assess a study's risk of bias.[13,14] Future methods guidance should consider addressing negative spillover, just as current guidance often discusses bias due to lack of blinding, differential attrition, or contamination.[2,4,13,14,63,67,68]

**Conclusion**

Interest in pragmatic trials continues to grow.[72,73] Artificial intelligence-based interventions increasingly shape clinical decision-making, which has led to calls for more randomized trials evaluating their impact on patient outcomes.[74-76] As new clinical practices and technologies that affect care delivery are considered for widespread adoption, well-designed trials are needed to provide valid evidence on their risks and benefits. Understanding all sources of bias that could affect these trials, including negative spillover, is a critical part of this effort.

Finally, some past trials—and systematic reviews, meta-analyses, or clinical practice guidelines that relied on their findings—may benefit from reassessment in light of this overlooked source of bias.

**Figure 1. Conditions That Lead to Negative Spillover**

There are four conditions that lead to negative spillover, each of which concerns a different portion of the trial pathway:
(1) Treatment and control arms draw from a shared resource pool (green dashed rectangle);
(2) The presence of a scarce resource in the trial setting (red dashed circle);
(3) The study intervention affects resource utilization (blue dashed circle); and
(4) Resource availability affects study outcomes (orange dashed circle).

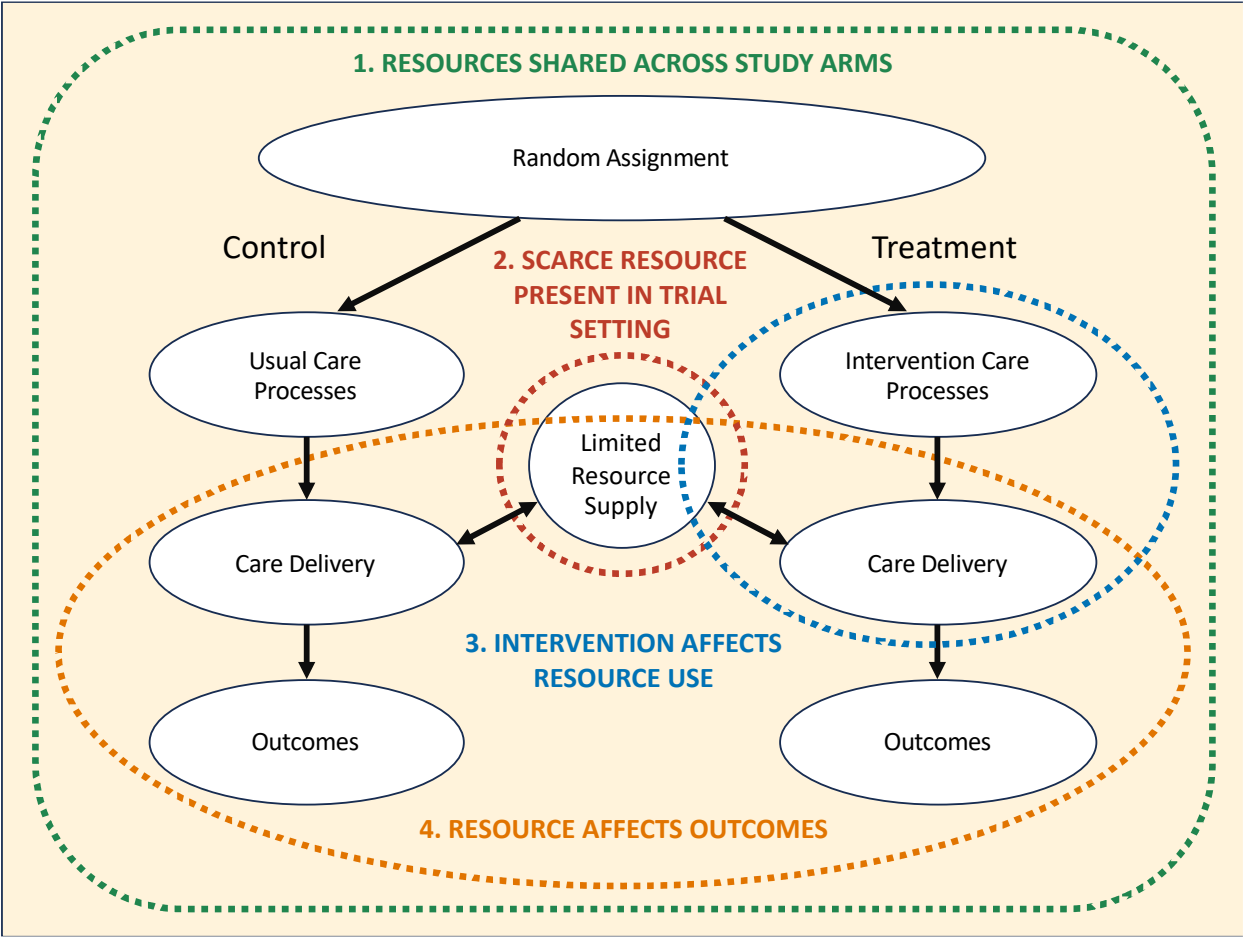

**Figure 2. Alternative Trial Designs that Mitigate Negative Spillover**

Two alternative individually randomized group-treatment[77] trial designs might be considered to mitigate negative spillover when multi-site cluster-randomization is not feasible. In a resource-balanced trial, a resource (such as individual appointment slots) could be assigned to the control or treatment group in the same proportion as individual patients. In a demand-balanced trial design, the intervention (such as a physiological alarm system) could be calibrated to generate the same average level of demand for care per patient as is generated by usual care processes.

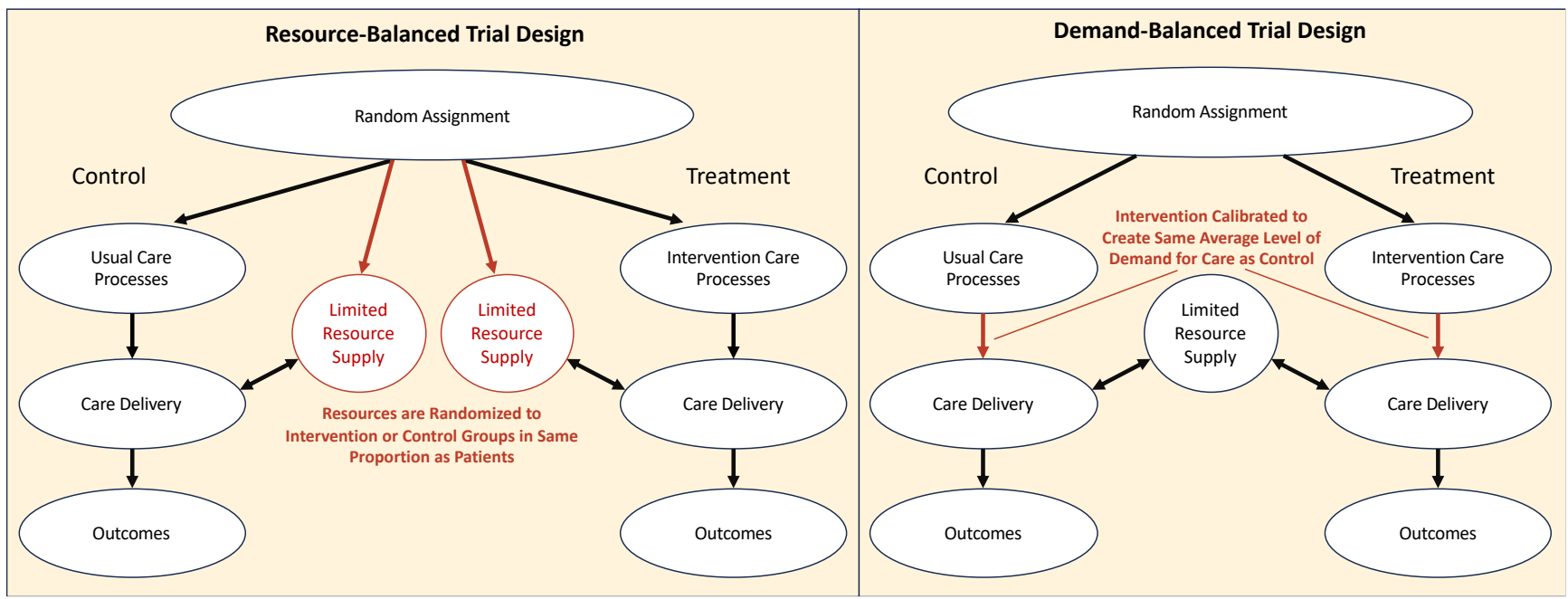

**Table 1. Trials in which Conditions Were Potentially Met for Negative Spillover***

| Study | 1. Resource Shared Across Study Arms | 2. Scarce Resource Present In Study Setting | 3. Study Intervention Affects Resource Use | 4. Study Outcome Affected By Resource Availability |
|---|---|---|---|---|
| Randomized trial of patient navigation for urban minority women with abnormal mammograms.[25] | Patients individually randomized to patient navigation or control groups, which received care at the same breast clinic. | Resource: Specialist appointments. Concerns over appointment availability and wait times for diagnostic services following abnormal cancer screening results are common.[21,22,32,33] | Navigators helped patients schedule appointments. The navigation group obtained care leading to diagnostic resolution earlier than control patients. | Outcomes: Time to diagnosis, patient anxiety, patient satisfaction. Increased demand for specialist appointments has been linked to increased wait times for cancer screening procedures.[32,33] Scheduling difficulties can delay breast and colorectal cancer diagnosis.[78-80] |
| Randomized trial of patient navigation in an underserved population.[12] | Patients individually randomized to patient navigation or control groups, which received care in the same health system. | Resource: Specialist appointments. Concerns over appointment availability and wait times for diagnostic services following abnormal cancer screening results are common.[21,22,32,33] | Navigators helped patients schedule appointments. The navigation group obtained care leading to diagnostic resolution earlier than control patients. | Outcome: Time to resolution of an abnormal screening finding. Increased demand for specialist appointments has been linked to increased wait times for cancer screening procedures.[32,33] Scheduling difficulties can delay breast and colorectal cancer diagnosis.[78-80] |
| Cluster randomized trial of EHR trigger-based reminder system to reduce diagnostic delays.[51] | Primary care providers randomized to intervention or control groups. Their patients received care in the same health systems. | Resource: Specialist appointments. Concerns over appointment availability and wait times for diagnostic services following abnormal cancer screening results are common.[21,22,32,33] | Intervention involved reminder calls to primary care providers to schedule patients for diagnostic services. The navigation group obtained care leading to diagnostic resolution earlier than control patients. | Outcome: Time to diagnostic evaluation. Increased demand for specialist appointments has been linked to increased wait times for cancer screening procedures.[32,33] Scheduling difficulties can delay breast and colorectal cancer diagnosis.[78-80] |
| Randomized Trial of Induction Versus Expectant Management (ARRIVE).[35,81] | Patients individually randomized to induction or expectant management groups, which received care in the same hospitals. | Resource: Clinicians; labor and delivery rooms. Delivery rooms unavailable for a few trial participants. Other studies of perinatal inpatient settings report lower than recommended nurse staffing levels[82] and that some nurses are routinely "overwhelmed."[83] | Elective induction group spent 45% longer in labor and delivery but had fewer cesarean deliveries and shorter postpartum stay. Some inductions rescheduled when rooms were unavailable. | Outcomes: "Composite of perinatal death or severe neonatal complications," cesarean delivery, duration of stay. Other studies report adverse birth outcomes increase on days with more patients or fewer clinicians[37] and that higher workloads may lead to higher risks to patients,[17] higher rates of cesarean delivery,[39] or lower rates of other procedures.[38] |
| Randomized trial of machine learning-based sepsis | Patients individually randomized to usual sepsis detection system or usual | Resource: Clinician time and attention. Concerns over clinician-to-patient ratios, staffing shortages, and | Patients monitored by the added physiological alarm system received earlier tests and treatment for sepsis. | Outcomes: duration of ICU stay, in-hospital mortality. ICU nurse availability, in terms of workforce size and workload, has been linked to |
| prediction algorithm.[42] | system plus alerts from an additional sepsis prediction algorithm. Patients received care in the same ICUs. | care rationing in ICU settings are common.[16,84-86] Clinicians' lack of time is cited as a barrier to providing patient care and following infection-control procedures in the ICU.[43-46,87] | No information provided on number of added alarms for intervention patients. Other studies have reported that responding to alarms adds to clinician workload and can disrupt patient care.[88-90] | patient mortality.[20,91] Exposure to excessive numbers of alarms can overload clinicians and lead to missed alarms and potentially patient deaths.[47] |
| Randomized trial of artificial intelligence-enabled electrocardiographic alarm system (AI-ECG).[92] | In-hospital patients individually randomized to usual care or monitoring by AI-ECG alarm system. Patients received care from same clinicians and hospital units. | Resource: Clinician time and attention; ICU beds. Concerns over clinician-to-patient ratios and staffing shortages in hospital and ICU settings are common.[16,84-86,91,93] ICU beds are not always available to meet the needs of all patients.[5,16,24,94] | Intervention effect "may be attributed to the increased attention of enrolled physicians." AI-ECG system monitored 8,001 patients and generated 709 alarms. Patients assigned to AI-ECG alarm system were more likely to receive medical interventions or be admitted to the ICU. | Outcomes: All-cause mortality, medical care utilization. Clinician availability and workload has been linked to in-hospital mortality.[19,91,93] Patients who are not admitted to an ICU due to lack of beds face higher mortality risks.[24,94] |
| Randomized trial of rapid whole-genome sequencing (NSIGHT1).[50] | Critically ill infants were individually randomized to receive standard genetic tests or standard tests plus a new genomic test. All tests were performed in same laboratory. | Resource: Laboratory staff and other resources. Turnaround of test results took multiple days. Others have raised concerns that staffing in clinical laboratories is sometimes inadequate.[95] | The intervention consisted of performing a new genomic test in addition to standard tests, which increases laboratory workload. NICU genetic sequencing and interpretation can take multiple hours or days of laboratory staff time.[96,97] | Outcome: Number of infants receiving genetic diagnosis within 28 days. Turnaround time for standard and additional genomic tests directly affects rate of diagnosis within 28 days. Concerns have been raised that inadequate staff resources can affect turnaround time in clinical laboratories.[95] Added workload may lead to delayed test results for all patients. |
| Randomized trial of text-messages to encourage uptake of COVID-19 vaccination.[49] | Patients invited for COVID-19 vaccination were randomized to receive text-message reminders or no reminders. Patients scheduled vaccine appointments in the same health system. | Resource: Vaccine doses and appointment slots. The trial began January 2021. The number of invited patients was limited to ensure sufficient vaccine supply, but appointment availability was not reported. Other vaccination programs in the same county and university system reported limited appointment availability for eligible adults at this time.[98,99] | Patients who received text-message reminders scheduled vaccination appointments earlier and at a higher rate. | Outcome: Number of participants scheduling a vaccination appointment within six days. Availability of vaccination appointments at convenient times could have affected whether an individual scheduled an appointment. |

* Most trials provided limited or no information on resource availability, utilization, or effect on outcomes. As a result, I use evidence from other studies in similar settings to discuss the potential role that a scarce resource may have played during the trial.